\documentstyle{mn}
\input epsf

\def\apj{ApJ}

\def\aj{AJ}
\def\aap{A\&\hskip-1pt A}

\def\mnras{MNRAS}

\def\lesssim{\mathrel{\hbox{\rlap{\hbox{\lower4pt\hbox{$\sim$}}}\hbox{$<$}}}}
\def\gtrsim{\mathrel{\hbox{\rlap{\hbox{\lower4pt\hbox{$\sim$}}}\hbox{$>$}}}}
\newcommand{\deltavec}{\mbox{\boldmath $\delta$}}
\newcommand{\zetavec}{\mbox{\boldmath $\zeta$}}
\newcommand{\rvec}{\mbox{\boldmath $r$}}
\newcommand{\uvec}{\mbox{\boldmath $u$}}

\newcommand{\zvec}{\mbox{\boldmath $z$}}

\title[Astrometric Multiple Lens Detection]
        {Astrometric Microlensing: A Channel to Detect 
         Multiple Lens Systems}

\author[Han]
{Cheongho Han\thanks{e-mails: cheongho@astroph.chungbuk.ac.kr (CH)}\\
Department of Physics, Institute for Basic Science Research, Chungbuk 
National University, Chongju 361-763, Korea}

\begin{document}
\maketitle
\vspace{-\abovedisplayskip}
\begin{abstract}
If a source star is gravitationally microlensed by a multiple lens system, 
the resulting light curve can have significant deviations from the standard 
form of a single lens event. The chance to produce significant deviations 
becomes important when the separations between the component lenses are 
equivalent to the combined angular Einstein ring radius of the system.  
For multiple lens systems composed of more than two lenses, however, this 
condition is difficult to meet because the orbits of such systems are 
unstable.  Even if events are caused by a multiple lens system with stable 
orbits where a pair of lenses are closely located and the other component 
(third body) has a wide separation from the pair, photometrically identifying 
the lens multiplicity will be difficult because the event will be identified 
either by a binary lens event caused by the close pair lenses or a single 
lens event caused by the third body.  In this paper, we show that if a 
seemingly binary lens event is astrometrically followed up by using future 
high precision interferometers, the existence of an additional third body 
can be identified via a repeating event. We show that the signatures of 
third bodies can be unambiguously identified from the characteristic
distortions they make in the centroid shift trajectories.  We also show 
that due to the long range astrometric effect of third bodies, the 
detection efficiency will be considerable even for third bodies with 
large separations from their close lens pairs.
\end{abstract}


\section{Introduction}

If a source star is gravitationally microlensed by a multiple lens system, 
the resulting light curve can have significant deviations from the standard 
form of a single lens event.  The chance to produce significant deviations 
induced by the lens multiplicity becomes important when the separations 
between the component lenses are comparable to the combined Einstein ring 
radius of the system, which is related to the physical parameters of 
the lens system by
\begin{equation}
\theta_{\rm E} = \left[{4G m_{\rm tot}\over c^2}
\left({1\over D_{\rm ol}}-{1\over D_{\rm os}} \right) \right]^{1/2},
\end{equation}
where $m_{\rm tot}=\sum_i^N m_i$ is the total mass of the lens system, $N$ 
is the total number of the component lenses, $m_i$ are the masses of the 
individual lenses, and $D_{\rm ol}$ and $D_{\rm os}$ are the distances 
to the lens system and the source star, respectively.  Some fraction of 
binary lens systems can meet this requirement, and dozens of candidate 
binary lens events have been reported by the lensing survey and followup 
observations (Udalski et al.\ 1994; Dominik \& Hirshfeld 1994; Alard, 
Mao \& Guibert 1995; Alcock et al.\ 2000; Afonso et al.\ 2000; Albrow et 
al.\ 2000).

However, for multiple lens systems that are composed of more than two 
lenses, it is difficult to photometrically identify the lens multiplicity.
There are two reasons for this.  First, multiple systems with all constituent 
lenses having separations between them equivalent to $\theta_{\rm E}$ are 
very rare because the orbits of such systems are unstable.  Second, even if 
events are caused by a multiple lens system with stable orbits where a pair 
of lenses are closely located and the other component (third body) has a 
wide separation from the pair, the close pair and the third body will 
behave as if they are two independent lens systems, causing the event to 
be identified either by a binary lens or a single lens event.  One case the 
lens multiplicity can be identified is when the source trajectory approaches
closely both the close lens pair and the third body, causing a repeating  
event.  However, due to the short range photometric effect of the third body,
the chance to produce repeating events is very low (see \S\ 3).  
Although there was a claim that a triple lens system was discovered by
Bennett et al.\ (1999), it was subsequently shown to be better explained
by a rotating binary (Albrow et al.\ 2000).  As a result, 
despite the substantial fraction of multiple systems (Evans 1968; Batten 
1973; Batten \& Fletcher 1989; Fekel 1981; Mayor \& Mazeh 1987), not a single 
multiple lens event has been reported to date.

Until now, lensing observations have been carried out only photometrically.
However, by using several planned high precision optical interferometers, 
such as those to be mounted on space-based platforms, e.g. the {\it Space 
Interferometry Mission} (SIM) and the {\it Global Astrometric Interferometer 
for Astrophysics} (GAIA), and those to be mounted on very large ground-based 
telescopes, e.g.\ Keck and VLT, it will become possible to observe lensing 
events astrometrically.  When an event is astrometrically observed by using 
these instruments, one can measure the lensing induced displacement of the 
source star image centroid position with respect to its unlensed position 
(centroid shift $\deltavec$).  Astrometric lensing observation is important 
because the lens mass can be better constrained with the measured centroid 
shift trajectory (Miyamoto \& Yoshii  1995; Walker 1995; Paczy\'nski 1998; 
Boden, Shao \& van Buren 1998).

In this paper, we show that astrometric lensing can also be used for 
efficiently detecting third bodies in multiple lens systems.  This is 
possible because due to the long range astrometric effect of third bodies, 
the lens multiplicity of a large fraction of events can be identified via 
repeating events.  We investigate the properties of the deviations induced 
by third bodies in the centroid shift trajectories of events and estimate 
the efficiency of third body detections expected from the future lensing 
observations.  We note that this paper is extension of the work of Han et 
al.\ (2002) who recently demonstrated the high efficiency of astrometric 
lensing observations in detecting very wide binary companions.

\section{Basics of Multiple Lensing}
If a source star is lensed by a multiple lens system, the locations of the 
resulting images are obtained by solving the lens equation.  When all lengths 
are normalized by the combined Einstein ring radius, the lens equation is 
expressed in complex notations by
\begin{equation}
\zeta=z+\sum_i^N {m_i/m_{\rm tot}\over \bar{z}_i - \bar{z}},
\end{equation}
where $z_i$ are the positions of the lenses, 
$\zeta=\xi+i\eta$ and $z=x+iy$ are the positions of the source and images, 
and $\bar{z}$ denotes the complex conjugate of $z$ (Witt 1990).  The 
magnifications of the individual images are given by the inverse of the 
determinant of the Jacobian of the lens equation evaluated at the image 
position, i.e.
\begin{equation}
A_j=\left( \frac{1}{|{\rm det}\ J|} \right)_{z=z_j};\ \ 
    {\rm det}\ J=1-{\partial \zeta \over \partial \bar{z}}
     {\overline{\partial \zeta} \over \partial \bar{z}}.
\end{equation}
Then the total magnification and the source star image centroid shift are 
obtained by
\begin{equation}
A=\sum_{j}^{N_I} A_j,
\end{equation}
and
\begin{equation}
\deltavec = {\sum_j^{N_I} A_j \zvec_j \over A} - \zetavec,
\end{equation}
where $\zetavec=(\xi,\eta)$ and $\zvec_j=(x_j,y_j)$ are the vector notations 
for the source and image positions and $N_I$ represents the number of images.  
Since the lens equation describes a mapping from the lens plane onto the 
source plane, to find the image positions for given positions of the source 
and the lenses, it is required to invert the lens equation.

For a single lens system ($N=1$), the lens equation can be simply inverted.
Algebraically solving the lens equation yields two solutions of the image 
positions, and the total magnification and the centroid shift are expressed 
in simple forms of
\begin{equation}
A = {u^2+2\over u \sqrt{u^2+4}},
\end{equation}
and
\begin{equation}
\deltavec = {\uvec \over u^2+ 2}\theta_{\rm E},
\end{equation}
where $\uvec$ is the dimensionless lens-source separation vector normalized
by $\theta_{\rm E}$.  For a single lens event, the light curve has a smooth
symmetric shape (Paczy\'nski 1986) and the centroid shift follows an 
elliptical trajectory (Walker 1995; Jeong, Han \& Park 1997).

For a multiple lens system ($N\geq 2$), on the other hand, the lens equation 
is non-linear and thus cannot be analytically inverted.  However, since the 
lens equation can be expressed as a polynomial in $z$, the image positions 
are obtained by numerically solving the polynomial.  For a $N$ point-mass 
lens system, the lens equation is equivalent to a $(N^2+1)$-order polynomial 
in $z$ and there are a maximum $N^2+1$ and a minimum $N+1$ images and the 
number of images changes by a multiple of two as the source crosses a caustic 
(Rhie 1997; Witt 1990).  The caustic is the main new feature of the multiple 
lens system, which refers to the source position on which the magnification 
of a point source event becomes infinity. For a binary lens system ($N=2$), 
the sets of caustics form close curves.  For a multiple lens system, the 
caustic structure becomes more complex and can exhibit self-intersecting and 
nested shapes.  The order of the polynomial rapidly increases as the number 
of lenses increases, and thus it becomes difficult to directly solve the 
equation for systems composed of many lenses.  One commonly used numerical 
method that allows one to study the lensing behaviors of multiple lens 
systems regardless of the number of lenses is the ``inverse ray-shooting 
method'' (Schneider \& Weiss 1986; Kayser, Refsdal \& Stabell 1986; Wambsganss 
1997).  The disadvantage of using this method is that it requires large 
computation time to study the detailed structures in the patterns of 
magnifications and centroid shifts.

In our analysis, we investigate the lensing properties of triple lens systems 
instead of testing systems with various numbers of lenses.  We note, however, 
that for systems composed of more than three lenses, the individual sets of 
lenses composed of the close pair lenses and each of the widely separated 
companion, in most cases, can be treated as independent triple lens systems.  
In addition, for triple lens systems, one can directly solve the lens equation 
instead of using the very time-consuming inverse ray-shooting method.

\section{Properties of Multiple Lensing Behaviors}

If a lens system contains an additional third body, the lensing behavior of 
the lens system is affected by the third body.  The most important effect of 
the third body is that it makes the effective positions of other lens 
components shifted towards it.  The approximate amount of the shift is 
\begin{equation}
\Delta z_i = {q_{\rm T}\over d}{\rvec_3\over \left\vert \rvec_3 \right\vert},
\end{equation}
where $\rvec_3$ is the position vector of the third body from the center of 
mass of the two close pair lenses (hereafter binary center), $d=\left\vert 
\rvec_3 \right\vert/ \theta_{\rm E,B}$ is the separation between the third 
body and the binary center normalized by the combined Einstein ring of the 
two close pair lenses, $\theta_{\rm E,B}={[4G(m_1+m_2)/c^2] [(1/D_{\rm ol})-
(1/D_{\rm os})]}^{1/2}$, and $q_{\rm T}=m_3/(m_1+m_2)$ is the ratio of the 
third body mass to the total mass of the close pair lenses.  If this effect 
is taken into consideration, however, the photometric lensing behavior of 
events during the time when the source passes the region around the close 
lens pair is well approximated by that of the binary lens events without the 
third body.  Similarly, besides the inverse effective positional shift of 
the third body towards the close lens pair, the lensing behavior of events
during the source's approach to the third body is well approximated by that 
of the single lens event caused solely by the third body.  That is, the close 
lens pair and the third body behave as if they are two independent lens 
systems.  In addition, since the photometric effect of both the close lens 
pair and the third body are confined to narrow regions around the individual 
lens systems, even if an event is caused by a triple lens systems, the event, 
in most cases, will be identified either by a binary lens or a single lens 
event.

However, if a seemingly binary lens event is astrometrically followed up, 
the existence of an additional third body can be identified with a 
significantly increased efficiency.  This is because, compared to the 
photometric effect, the astrometric effect of the third body endures to 
a large distance from it (Miralda-Escud\'e 1996), and thus it can cause 
noticeable deviations in the trajectories of the source star image centroid 
shifts even when the source approaches the third body with a considerable 
separations.

\begin{figure}
\vskip-0.4cm
\epsfysize=12.0cm
\centerline{\epsfbox{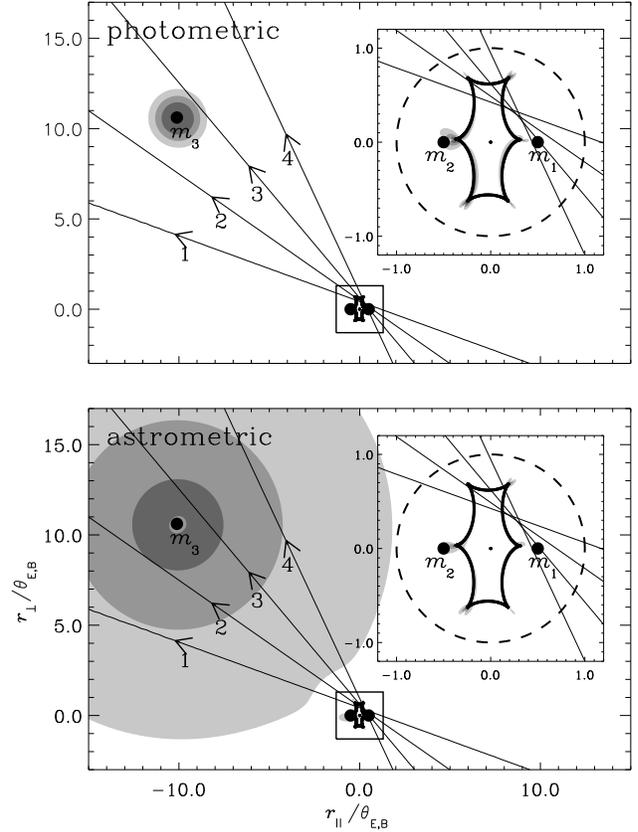}}
\vskip-0.5cm
\caption{
The maps of excess magnification (upper panel) and centroid shift (lower panel) 
of an example multiple lens system.  The lens system is composed of three 
lenses where a pair of lenses are closely located and the other component 
(third body) is widely separated from the pair.  The two close pair lenses 
have a mass ratio of $q_{\rm B}=m_2/m_1=1.0$ and they are separated by 
$a_{\rm B}=1.0$ in units of $\theta_{\rm E,B}$.  The third body has a mass 
ratio of $q_{\rm T}=m_3/(m_1+m_2)=0.6$ and separated from the center of mass 
of the close pair lenses by $a_{\rm T}=15.0$ also in units of $\theta_{\rm E,B}$.
The filled dots represent the locations of the three lenses.  Grey scales are 
used to represent the regions of deviations with $\epsilon \geq 5\%$, 10\%, 
and 20\% for the excess magnification map and $\Delta\delta \geq 5\%$, 10\%, 
and 20\% of $\theta_{\rm E,B}$ for the excess centroid shift map.  The coordinates 
of the maps are centered at the center of the mass of the close pair lenses and
all lengths are normalized by $\theta_{\rm E,B}$.  The straight lines with
arrows represent the source trajectories of events whose light curves and
centroid shifts are presented in Fig.\ 2.  The solid curve near the close
lens pair represents the caustics.  The inset in each panel shows the enlarged
view of the map in the region around the close pair lenses.  The dashed circle 
in each inset represents the combined Einstein ring of the close pair lenses.
}
\end{figure}

\begin{figure*}
\vskip-1.2cm
\epsfysize=16cm
\centerline{\epsfbox{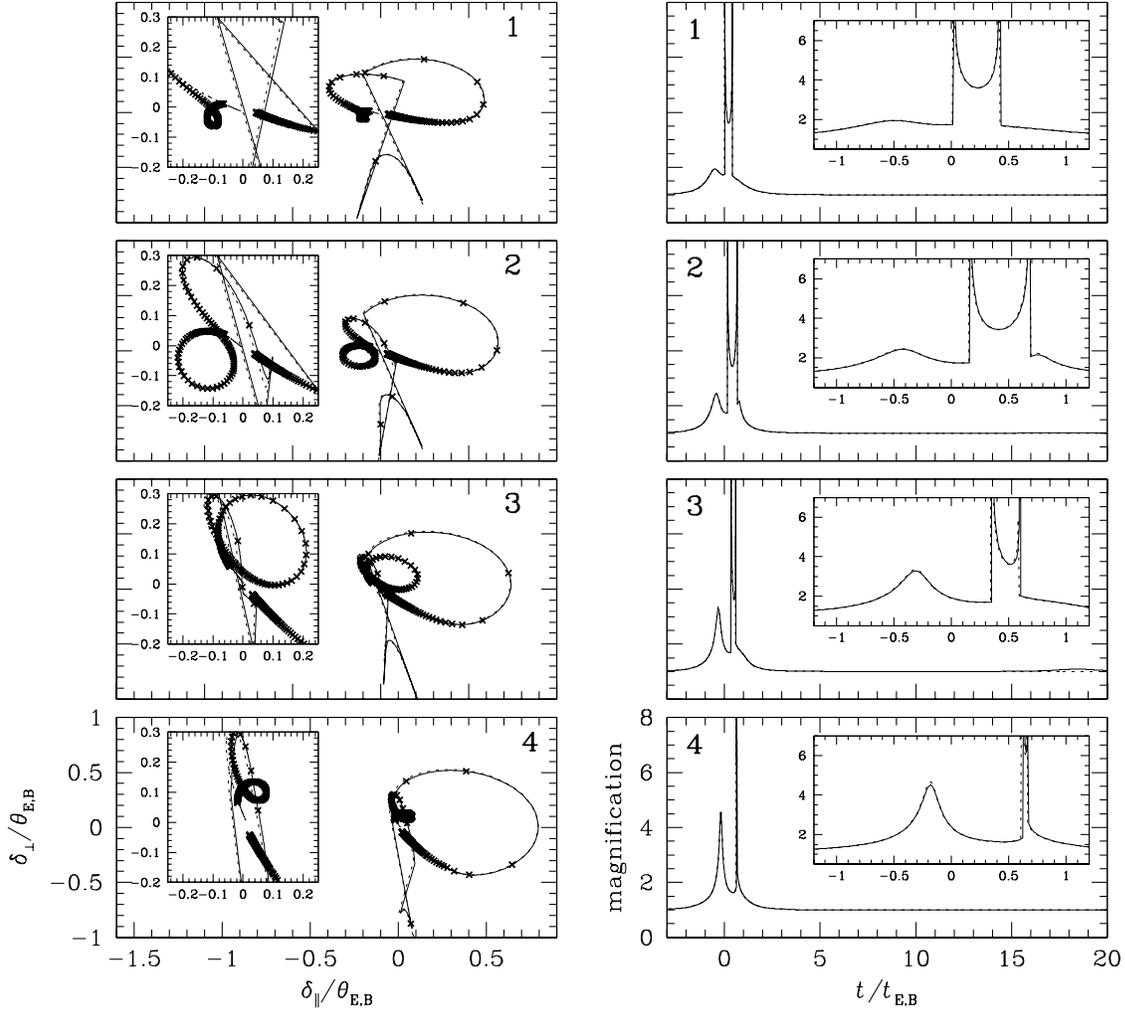}}
\vskip-1.2cm
\caption{
Example centroid shift trajectories (left panels) and light curves (right
panels) of events caused by the triple lens system whose maps of excess
magnification and centroid shift are presented in Fig.\ 1.  The source
trajectories responsible for the individual events are marked in Fig.\ 1,
where the number in each panel corresponds to that of the source trajectory.
In each panel, the dotted and solid curves represent the centroid shift
trajectories and light curves produced with and without the third body.  The
insets inside the left side panels show the detailed structures of the
centroid shift distortions induced by the third body.  The insets inside the
right side panels show the enlargement of the part of the light curve during
the source's passages around the region of the close lens pair.
}
\end{figure*}

To show this, we construct maps of excess magnifications and centroid shifts 
of an example triple lens system.  The excess magnification and the centroid 
shift are defined respectively by
\begin{equation}
\epsilon = {A_{\rm T}-A_{\rm B}\over A_{\rm T}},
\end{equation}
and
\begin{equation}
\Delta\deltavec = \deltavec_{\rm T}-\deltavec_{\rm B},
\end{equation}
where $A_{\rm T}$ and $\deltavec_{\rm T}$ represent the magnification and 
the centroid shift of the exact triple lens system, while $A_{\rm B}$ and 
$\deltavec_{\rm B}$ represent those of the binary lens event without the 
third body.  The constructed maps are presented in Figure 1.  For the 
construction of the maps, we take the effect of the positional shifts 
induced by the third body into consideration.  The two close pair lenses of 
the triple lens system have a mass ratio of $q_{\rm B}=m_2/m_1=1.0$ and the 
separation between them is $a_{\rm B}=1.0$ in units of $\theta_{\rm E,B}$.  
The third body has a mass ratio of $q_{\rm T} =0.6$ and it is separated from 
the binary center by $a_{\rm T}=15.0$ (also in units of 
$\theta_{\rm E,B}$\footnote{According to the strategy of monitoring binary 
lens events for third body detections, the event caused by the close 
pair lenses will be the standard of all measurements.  Therefore, we normalize 
all lengths and time scales in units of $\theta_{\rm E,B}$ and $t_{\rm E,B}$, 
where $t_{\rm E,B}$ is the time required for the source to transit 
$\theta_{\rm E,B}$.}) with an orientation angle of $45^\circ$ with 
respect to the axis connecting the two close pair lenses.  The locations of 
the lenses are marked by filled dots on the maps.  Grey scales are used to 
represent the regions of deviations with $\epsilon \geq 5\%$, 10\%, and 20\% 
for the excess magnification map and $\Delta\delta \geq 5\%$, 10\%, and 20\% 
of $\theta_{\rm E,B}$ for the excess centroid shift map.  For a typical 
Galactic lensing event, the angular Einstein ring radius is $\sim ({\cal O}) 
10^2$ $\mu$-arcsec, and thus $\Delta\delta=0.1\theta_{\rm E,B}$ corresponds 
to $\sim ({\cal O})10$ $\mu$-arcsec.  For comparison, we note the SIM will be 
able to measure positional shifts of stars as small as several 
$\mu$-{\rm arcsec (http://sim.jpl.nasa.gov).

From Fig.\ 1, one finds the following facts.  First, due to the large 
separation between the third body and the close lens pair, both the 
photometric and astrometric effects of the third body are not important in 
the region around the close lens pair, implying that not only the photometric 
but also the astrometric lensing behaviors of triple lens events during the 
source's passage of the region around the close lens pair are well described 
by the binary lensing approximation.  Second, the region of significant 
astrometric deviations around the third body is much larger than the region 
of significant photometric deviations, implying that for an important fraction 
of multiple lens events the existence of third bodies can be identified from 
extended astrometric followup lensing observations.

To see the characteristics of the photometric and astrometric deviations 
induced by the third body, in Figure 2, we present the light curves and the 
centroid shift trajectories of several events caused by the triple lens 
system.  The source trajectories responsible for the individual events are 
marked in Fig.\ 1.  In each panel, we also present the curves of the binary 
lens events computed without the third body taking the effective positional 
shifts of the close binary lens pair into consideration (dotted curves).  
On the centroid shift trajectories, we mark the centroid positions (`{\tt x}' 
symbols) measured with a time interval of $t_{\rm E,B}/4$ to show the changing 
rate of the centroid position.  For an event with $t_{\rm E,B}\sim 1$ month,
therefore, this interval corresponds to roughly a week.  From the comparison 
of the centroid shift trajectories and the light curves of the corresponding 
events, one finds that the astrometric signatures of the third body can be 
clearly identified from the characteristic loop-shaped distortions in the 
centroid shift trajectories, while the photometric signatures are too week 
to be noticed.  One also finds that the astrometric deviations last for a 
long period of time.

\section{Efficiency of Third Detections}
In the previous section, we showed that astrometric lensing observations 
will enable one to identify the third body of a multiple lens system with 
little ambiguity from the characteristic deviations it makes in the centroid 
shift trajectories.  In this section, we determine the efficiency of detecting 
third bodies expected from the future astrometric lensing observations.  For 
this determination, we use the formalism introduced by Di Stefano \& Scalzo 
(1999) and further developed by Han et al.\ (2002).

If we define $b_{\rm B}$ and $b_3$ as the smallest separations from the 
source trajectory to the binary center and the third body, the orientation 
angle of the source trajectory with respect to the line connecting the 
binary center and the third body is represented by
\begin{equation}
\alpha_\pm = \sin^{-1}\left( {b_{\rm B}\pm b_3 \over 
\left\vert \rvec_3\right\vert}\right),
\end{equation}
where the sign `$-$' is for the case when the closest points on the source 
trajectory from the binary center and the third body are on the same side 
with respect to the line connecting the binary center and the third body,
and the `$+$' sign is when the closest points are on the opposite sides.
If all lengths are normalized in units of $\theta_{\rm E,B}$, equation 
(11) is expressed by
\begin{equation}
\alpha_\pm = \sin^{-1}\left( {\beta_{\rm B}\pm \sqrt{q_{\rm T}}\beta_3 
\over d}\right),
\end{equation}
where $\beta_{\rm B}$ and $\beta_3$ are the lensing impact parameters of the 
two independent events involved with the binary system composed of the close 
pair lenses and the single third body, respectively.\footnote{By definition, 
the lensing impact parameter represents the closest separation between the 
source trajectory to the lens (the center of mass of the lens system) 
normalized by the (combined) angular Einstein ring radius of the involved 
lens (lens system), and thus $\beta_{\rm B}=b_{\rm B}/\theta_{\rm E,B}$ and 
$\beta_3=b_3/\theta_{\rm E,3}$, where $\theta_{\rm E,3}$ represents the 
Einstein ring radius of the third body.} Let us additionally define 
$\beta_{3,{\rm th}}$ such that among events that were identified to be 
affected by the close lens pair by approaching the binary center with an 
impact parameter $\beta_{\rm B}$, only events with source trajectories 
passing the third body closer than $\beta_{3,{\rm th}}$ can be identified 
as multiple lens events from the centroid shift deviations larger than a 
threshold value $\Delta\delta_{\rm th}$.  Then, the fraction of multiple lens 
events whose third bodies can be astrometrically identified is computed by
\begin{equation}
P = {\alpha_{+,{\rm th}}-\alpha_{-,{\rm th}} \over \pi},
\end{equation}
where the threshold orientation angles have values
\begin{equation}
\cases{
\alpha_{+,{\rm th}}={\rm sin}^{-1}\hskip-1pt \left({\beta_{\rm B}+
\sqrt{q_{\rm T}}\beta_{3,{\rm th}}\over d} \right) \cr
\alpha_{-,{\rm th}}={\rm sin}^{-1}\hskip-1pt \left({\beta_{\rm B}-
\sqrt{q_{\rm T}}\beta_{3,{\rm th}}\over d} \right) \cr
}
\end{equation}
 for $d> \beta_{\rm B}+\sqrt{q_{\rm T}}\beta_{3,{\rm th}}$,
\begin{equation}
\cases{
\alpha_{+,{\rm th}}= \pi/2 \cr
\alpha_{-,{\rm th}}={\rm sin}^{-1}\hskip-1pt \left({\beta_{\rm B}-
\sqrt{q_{\rm T}}\beta_{3,{\rm th}}\over d} \right) \cr
}
\end{equation}
for $\left\vert\beta_{\rm B}-\sqrt{q_{\rm T}}\beta_{3,{\rm th}}
\right\vert< d \leq \beta_{\rm B}+\sqrt{q_{\rm T}}\beta_{3,{\rm th}}$, and
\begin{equation}
\cases{
\alpha_{+,{\rm th}}= \pi/2 \cr
\alpha_{-,{\rm th}}=-\pi/2, \cr
}
\end{equation}
for $d \leq |\beta_{\rm B} - \sqrt{q_{\rm T}} \beta_{3,{\rm th}}|$.  We 
note that the threshold orientation angles have different values depending 
on the relative size of the astrometrically effective region of the third 
body compared to the separation between the third body and the binary 
center.  For a given value of the threshold centroid shift deviation, 
$\Delta\delta_{\rm th}$, the corresponding threshold impact parameter to 
the third body is obtained by
\begin{equation}
\beta_{{3,\rm th}} = {1\over 2}\left( {1\over \Delta\delta_{\rm th}/
\theta_{{\rm E},3}} + \sqrt{{1\over (\Delta\delta_{\rm th}/
\theta_{{\rm E},3})^2} - 8} \right).
\end{equation}
If one expresses the threshold astrometric deviation in terms of the fraction 
of the combined Einstein ring radius of the close pair lenses, 
$f=\Delta\delta_{\rm th}/\theta_{\rm E,B}$, equation (17) becomes 
\begin{equation}
\beta_{{3,\rm th}} = {1\over 2}\left( {\sqrt{q_{\rm T}}\over f}+
\sqrt{ {q_{\rm T}\over f^2}-8}\right).
\end{equation}

In Figure 3, we present the determined efficiency in the parameter space 
of the separation and the mass ratio between the third body and the close 
lens pair, $P(d,q_{\rm T})$.  For this computation, we set $f=0.1$.  For the 
impact parameter to the close lens pair, we adopt $\beta_{\rm B}=0.0$, but 
we note that the dependency of the efficiency on the adopted value of 
$\beta_{\rm B}$ is not important as long as $\beta_{\rm B}\ll d$.  We also 
note that $q_{\rm T}$ can be larger than 1.0 because the third body can be 
heavier than the total mass of the close pair lenses.  From the figure, one 
finds that the efficiency is substantial even for third bodies that are 
widely separated from their close lens pairs.  It is worth to note that the 
scaling between $d$ and $q_{\rm T}$ are quite linear.  This can be understood 
as follows.  The probability of detecting third bodies is proportional to the
threshold orientation angle, i.e.\ $P\propto \alpha \propto {\rm sin}^{-1}
(\sqrt{q_{\rm T}}\beta_{\rm 3,th}/d)$ [see eq.\ (12) and (13)].  In the limit 
of $d\gg \sqrt{q_{\rm T}}\beta_{\rm 3,th}$, ${\rm sin}^{-1}(\sqrt{q_{\rm T}}
\beta_{\rm 3,th}/d)\sim \sqrt{q_{\rm T}} \beta_{\rm 3,th}/d$, and thus 
$P\propto \sqrt{q_{\rm T}}\beta_{\rm 3,th}/d$.  Since $\beta_{\rm 3,th}\propto 
\sqrt{q_{\rm T}}$ [see eq.\ (18)], one finds that $P\propto q_{\rm T}/d$.  
Therefore, the scaling between $d$ and $q_{\rm T}$ is linear.

\begin{figure}
\vskip-0.6cm
\epsfysize=9.5cm 
\centerline{\epsfbox{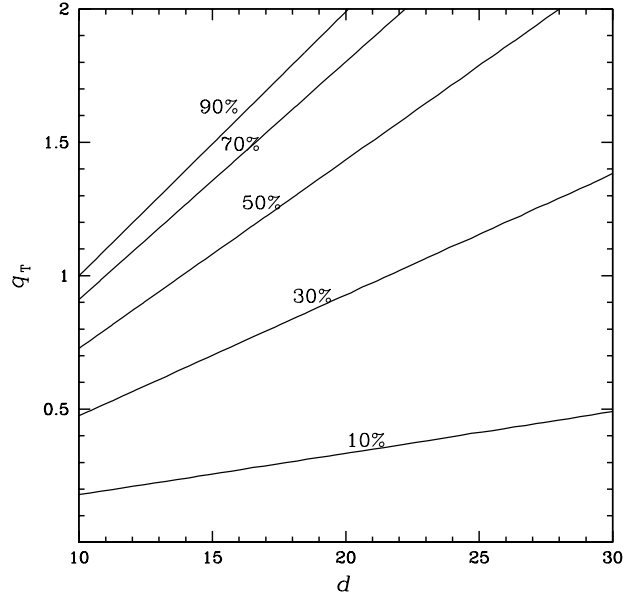}} 
\vskip-0.7cm
\caption{
The expected efficiency of detecting third bodies from astrometric followup 
lensing observations as functions of the separation, $d$, and the mass ratio,
$q_{\rm T}$, between the third body and the close lens pair.  We assume that 
a third body is detected if the deviation induced by the third body is larger 
than 10\% of the combined angular Einstein ring radius of the monitored binary 
lens event, $\theta_{\rm E,B}$.  The separation is expressed in units of 
$\theta_{\rm E,B}$.
}
\end{figure}

\section{Conclusion}

In this paper, we show that future astrometric lensing observations will 
provide an efficient method of  detecting third bodies of multiple lens 
systems, which could not have been detected from conventional photometric 
lensing observations.  We showed that the deviation in the centroid shift 
trajectory induced by the third body of a multiple lens system has a 
characteristic loop that can be clearly distinguished from other types of 
deviations and thus can be unambiguously identified.  In addition, since 
the deviations last for a long period of time, detecting third bodies will 
be possible even from sparse astrometric sampling.

This work was supported by a grant (BSRI-01-2) from the Basic Science 
Research Institute of Chungbuk National University.

{}

\end{document}